\documentclass[graybox]{svmult}

\usepackage{mathptmx}       
\usepackage{helvet}         
\usepackage{courier}        
\usepackage{type1cm}        
\usepackage{makeidx}         
\usepackage{graphicx}        
\usepackage{multicol}        
\usepackage[bottom]{footmisc}

\makeindex             

\begin{document}

\title*{Cherenkov Telescope results on gamma-ray binaries}
\author{Juan Cortina}
\institute{Juan Cortina \at Institut de Fisica d'Altes Energies (IFAE), Edifici CN, 
Campus UAB, Cerdanyola del Valles, E-08193, Spain, \email{cortina@ifae.es}
}
%
%
\maketitle

\abstract*{
In the past ten years of regular operations, a new generation of Cherenkov telescopes 
have established binary systems as a new class of Very High Energy $\gamma$-ray (VHE) emitters.
Particle acceleration in these systems may occur either in an accretion-powered jet
(``microquasar'') or in the shock between a pulsar wind and a stellar wind (``wind-wind'').
This paper describes the phenomenology of the three VHE binaries PSR~B1259-63,
LS 5039 and LS I +61 303. Two other objects may belong to this new class:
HESS J0632+057 is a point-like variable VHE source whose multiwavelength behaviour
resembles that of the other binaries, whereas Cyg X-1 is a well-known accreting system
which may have been detected in VHE during a flaring episode. The paper concludes with a review
of the latest searches for other binaries with Cherenkov telescopes, with special 
emphasis on Cyg X-3.
}

\abstract{
In the past ten years of regular operations, a new generation of Cherenkov telescopes 
have established binary systems as a new class of Very High Energy $\gamma$-ray (VHE) emitters.
Particle acceleration in these systems may occur either in an accretion-powered jet
(``microquasar'') or in the shock between a pulsar wind and a stellar wind (``wind-wind'').
This paper describes the phenomenology of the three VHE binaries PSR~B1259-63,
LS 5039 and LS I +61 303. Two other objects may belong to this new class:
HESS J0632+057 is a point-like variable VHE source whose multiwavelength behaviour
resembles that of the other binaries, whereas Cyg X-1 is a well-known accreting system
which may have been detected in VHE during a flaring episode. The paper concludes with a review
of the latest searches for other binaries with Cherenkov telescopes, with special 
emphasis on Cyg X-3.
}

\section{Introduction}
\label{sec:0}

The VHE (or TeV) band covers photon energies in excess of a few tens of GeV and it is mainly studied 
with ground-based Imaging Atmospheric Cherenkov Telescopes (IACTs).
Results presented here mostly refer to the latest generation instruments H.E.S.S., MAGIC and VERITAS.
A recent review of this young field of astronomy can be found here\cite{vhe_review}.
High Energy $\gamma$-ray (HE) detectors on board satellites such as CGRO/EGRET, {\it AGILE} and {\it Fermi}/LAT 
are sensitive to photon energies from tens to MeV up to tens of GeV and are reviewed
elsewhere in this conference\cite{he_review}.

We will not dwell in this paper on the physics interpretation of the observational results, 
but refer the reader to other contributions to this 
conference\cite{observational_review,theoretical_review,cerutti,moldon,zabalza,casares2} 
and references within. Let us shortly mention however that
theoretical models fall into two possible scenarios for the production of VHE 
$\gamma$-rays in binary systems. In the ``microquasar'' scenario, particle acceleration takes place 
at a jet which originates at an accretion disk. This scaled-down version
of an active galactic nucleus may provide significant insights into the
mechanism of jet production and particle acceleration since all processes
take place at significantly shorter time scales. In the ``wind-wind'' scenario,
particle acceleration happens at the interaction region between a pulsar
wind and the wind of the companion star.

We will start by describing the VHE results on the three well established $\gamma$-ray 
binaries PSR~B1259-63, LS 5039, LS I +61 303. The compact object in the first 
system is a pulsar and the VHE emission can be well understood within the wind-wind scenario.
The physical interpretation for the other two systems is still controversial. 
We will then deal with the X-ray binary Cyg X-1, for which the only 
evidence for VHE emission comes from a short flare marginally detected by MAGIC, and
with another object, HESS J0632+057, which is an established source of VHE emission, 
but whose association to a binary system is still uncertain. We will close the paper with 
a review of several X-ray binaries which have been actively searched for at
VHE energies, but which remain undetected.

\section{Detected in VHE: PSR~B1259-63, LS 5039 and LS I +61 303}
\label{sec:1}

\subsection{PSR~B1259-63/LS~2883}

PSR~B1259-63/LS~2883 was the first binary established at VHE. 
It was discovered using the H.E.S.S. telescope array in 2004\cite{hess_1259}. 

The binary system is formed by a 48 ms pulsar and a B2Ve star at a distance of 
1.5 kpc (for a summary of the system parameters, see \cite{tavani}). There is no evidence 
for jets and the compact object is a pulsar,
so particle acceleration probably takes place at a shock between the pulsar wind and
the wind of the stellar companion. In fact VHE emission through Inverse Compton of 
shock-accelerated leptons had been predicted well before the detection\cite{kirk} and
PSR~B1259-63/LS~2833 had already been classified as a binary system with a plerionic 
component in \cite{tavani}.

The orbit is highly eccentric (e=0.87) and has a period of 3.4 years. Periastron takes
place at a distance of 0.7~A.U., while apastron happens at around 10 A.U. 
Close to periastron, the pulsar travels through the starʼs 
circumstellar disk, which is inclined 10-40$^{\circ}$ to the orbital plane. The 
radio pulse vanishes from around 15 days before to around 15 days 
after periastron due to absorption in the disk.

\begin{figure}[h]
\sidecaption
\includegraphics[scale=.30]{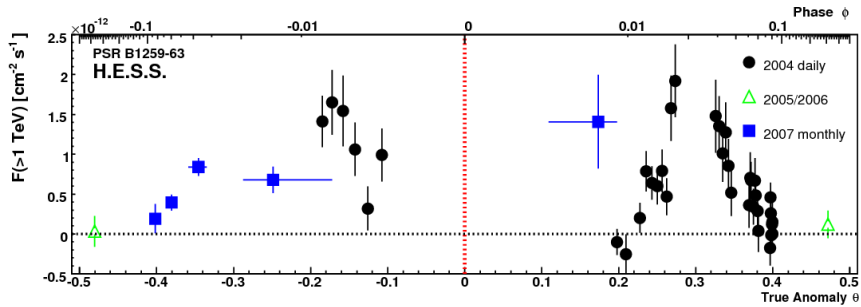}
%
%
\caption{VHE integrated flux from PSR~B1259-63 above 1 TeV as a function of the true anomaly.
The corresponding orbital phases (mean anomaly) 
are shown on the upper horizontal axis. The red vertical line indicates the periastron passage. 
Shown are data from the years 2004 to 2007: the 
black points are the daily fluxes as measured in 2004. The green empty triangles show the overall 
flux level as seen in 2005 and 2006. The blue filled 
squares represent the monthly fluxes taken from the campaign in 2007. 
From \cite{hess_1259_2nd_periastron}.}
\label{fig:1259_lightcurve}       
\end{figure}
The long and eccentric orbit complicates the VHE observations, as the source is
only bright in VHE for a few months every 3.4 years and observational constraints 
further limit the coverage during this period. In fact
the source could never be observed in the 2-3 days around periastron so far.
The first H.E.S.S. campaign during the periastron passage in 2004 revealed a 
complex light curve (see black points in Fig. \ref{fig:1259_lightcurve}). 
Observations could only start when the source was already 10 days
before periastron. The VHE flux was actually decreasing from a maximum of $\sim$10\% 
of the Crab Nebula flux on the first nights of observation. It followed a gap during 
periastron and the source was found to be brightening again up to another maximum 
at $\sim$10\% of Crab. The flux
then decreased steadily for the following three months. 
This peculiar behaviour has been linked to the influence of the Be star disc 
on the emission process\cite{1259_disk1,1259_disk2}.

The spectrum could be fitted to
a power law with $\Gamma$=2.7$\pm$0.2 (stat) $\pm$0.2 (sys) with no indication
of index variability. The VHE flux corresponds to a luminosity of 8$\times$10$^{32}$ erg/s, 
which represents only 0.1\% of the pulsar's spindown luminosity.

Figure \ref{fig:1259_lightcurve} shows the VHE flux of the source as a function of 
the true anomaly for both the 2004 periastron passage and the next passage in 
2007\cite{hess_1259_2nd_periastron}. It was again impossible to observe strictly 
during periastron. In general terms the source displays the same level of VHE emission,
but the two light curves show different shapes, even if there are
no observations at exactly the same true anomalies: the VHE seems to be brighter
at $\theta$=$-$0.3 in 2004, but dimmer after periastron, at $\theta$=+0.2. 
This different behaviour in 2007 challenges models where the double-hump
structure is associated to the pulsar crossing the disc.

\subsection{LS 5039}

This binary system consists of a compact object and an O6.5V star of 23~M$_{\odot}$. 
It is located a distance of $\sim$2.5~kpc. The orbit has a small 
eccentricity e=0.337 with a period of 3.9061$\pm$0.0001 days. 
The two objects are 0.1~A.U. away at periastron ($\phi$=0) and 0.2~A.U away
at apastron. More details about the latest orbital parameters and phase definition 
can be found at \cite{orbital_parameters}

\begin{figure}[h]
\sidecaption
\includegraphics[scale=.70]{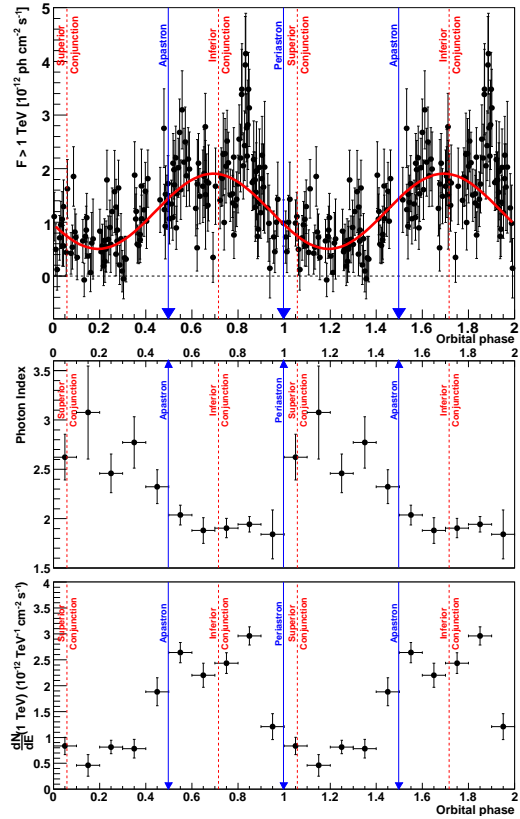}
%
%
\caption{Top: Integral $\gamma$-ray flux ($F>1$~TeV) lightcurve (phaseogram) 
of LS~5039 from H.E.S.S. data (2004 to 2005) on a 28 minute run basis folded with the orbital 
ephemeris in \cite{Casares}. The blue solid arrows correspond to periastron and apastron.
The thin red dashed lines represent the superior and inferior conjunctions of the compact object. 
Middle: Fitted pure power-law photon index (for energies 0.2 to 5~TeV) vs. phase 
interval of width $\Delta \phi=0.1$. Bottom: Power-law normalisation (at 1~TeV) vs. 
phase interval of width  $\Delta \phi=0.1$. From \cite{LS5039_modulation}.}
\label{fig:LS5039_modulation}       
\end{figure}
The nature of the compact object is unknown: it may range from a 1.4~M$_{\odot}$ neutron 
star to a 3.7~M$_{\odot}$ black hole. No pulsar has been found in radio or X-ray searches, 
although any pulsations would probably be diluted by Compton scattering for all 
orbital phases. As mentioned above, no pulsations are observed when the two components of the 
PSR~B1259-63/LS~2833 system approach periastron and the two components of LS 5039
are a comparable distance for all orbital phases.
VLBA shows complex extended morphology which may look at first sight like a jet, but
changes orientation as the two objects progress along the orbit\cite{LS5039_VLBI}. 
This argues in favor of acceleration at the region where the pulsar wind and the 
companion star's wind interact, rather than acceleration in a jet.

LS 5039 was discovered in VHE by H.E.S.S. \cite{LS5039} during their first Galactic Plane 
Survey. In fact it was the only point-like source found in the survey and its position
was consistent with a bright unidentified EGRET source which had been proposed 
as a counterpart to a binary system\cite{LS5039_egret}. The spectrum could be fitted
to a power law with photon index 2.12$\pm$0.15 and a flux of $\sim$5\% Crab. 
The VHE luminosity was similar to the X-ray luminosity.

Further H.E.S.S. observations in 2005 allowed to establish variability 
and periodicity consistent with the orbital period\cite{LS5039_modulation}. The 
phase-folded flux of the source is shown in figure \ref{fig:LS5039_modulation}
along with the evolution of the photon index with phase.
The VHE flux changes from $\sim$15\% Crab at inferior conjunction (when the compact 
object is between the companion star and the Earth) to $\sim$5\% Crab 
at superior conjunction and the spectral shape is strongly 
modulated along the orbit. In fact the spectrum at inferior conjunction is not a simple
power law and shows a clear hardening in the region 0.3 to $\sim$20 TeV.

\subsection{LS I +61 303}

LS I +61 303 is a binary system formed by a compact object and a B0Ve star of 
12M$_{\odot}$ at a distance of 2.0$\pm$0.2kpc. It has an eccentricity e=0.54 with 
26.4960 $\pm$ 0.0028 day period (from radio observations). At periastron the two system 
components are separated by 0.2 A.U. ($\phi$= 0.275, see \cite{orbital_parameters}
for the orbital elements of the system and the definition of the phase), while
at apastron they are 1 A.U apart. 

\begin{figure}[h]
\sidecaption
\includegraphics[scale=.50]{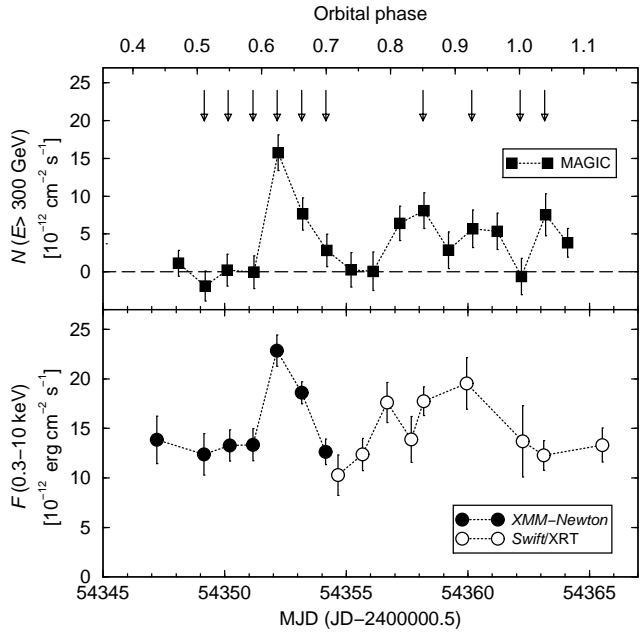}
%
%
\caption{VHE and X-ray light curves of LS I +61 303 during the multiwavelength
campaign of 2007 September. Top: MAGIC VHE flux above 300~GeV vs. the observation
time in MJD and the orbital phase. The horizontal dashed line indicates 0 flux.
The vertical arrows mark the times of simultaneous VHE gamma-ray and X-ray
observations. Bottom: de-absorbed flux in the 0.3--10~keV energy range
for the seven XMM observations (filled circles) and the nine {\it Swift}  ones
(open circles). Error bars correspond to a 1$\sigma$ confidence level in all
cases. Dotted lines join consecutive data points to help following the main
trends of the light curves. The sizes of the symbols are larger than the time
span of individual observations. From \cite{lsi_xray_correlation}.}
\label{fig:lsi_xray_vhe}       
\end{figure}
Historically this object has drawn much interest due to its periodic outbursts
in radio and X-rays. The radio outbursts are well correlated with the
orbital period\cite{gregory_radio_period}, 
although the phase of maximum emission moves from $\phi$=0.45 to
0.95 over a period of 1667 days. This superorbital variability is probably 
related to changes in the mass loss rate of the Be star and/or density 
of the circumstellar wind\cite{lsi_superorbital}. 

Even if extensively studied, we do not know the nature of the compact object: it could be 
anything from a 1.4M$_{\odot}$ neutron star to a 4M$_{\odot}$ black hole. No pulsar 
has been found in radio or X-ray searches, but the aforementioned considerations
about absorption for LS 5039 apply here as well. Also similarly to LS 5039, 
VLBA observations reveal a complex morphology which is coupled with 
the orbital period\cite{dhawan}, once again pointing to wind-wind interaction.

This binary system was discovered in VHE $\gamma$-rays by MAGIC in 2006\cite{lsi_discovery}
after following the source for a good fraction of the orbital phases over six orbital periods.
The VHE emission was significantly variable: there was no detection at periastron
or at inferior conjunction (i.e. when the compact object is between the companion star
and the Earth, the phase for which LS 5039 reaches a maximum); the emission peaks 
at $\sim$15\% Crab before apastron ($\phi$=0.6-0.7). 
The spectrum is Crab-like, with a photon index of 2.6 $\pm$ 0.2 (stat) $\pm$ 0.2 (syst).
The source was confirmed by the VERITAS telescope array during the next season 
\cite{lsi_veritas_confirmation}. The light curve and spectrum were found to 
be consistent with MAGIC results.

A second MAGIC campaign in 2006\cite{lsi_periodicity}
combined with previous 2005 data allowed to establish that the VHE emission 
has a significant periodic component with a period of 26.8$\pm$0.2 days, 
compatible with the orbital period and the period found in other wavelengths.
Contrary to what happens in LS 5039, the phaseogram follows no simple sinusoidal 
shape.

MAGIC observations showed evidence for a second peak of VHE emission in December 2006 
at $\phi$=0.8-0.9 but for only one orbit. No evidence was found for intranight 
variability in searches down to 15 min time scales, or for spectral variability, although 
it must be said that the spectral index could only be measured for phases $\phi$=0.5-0.6 
and $\phi$=0.6-0.7\cite{lsi_periodicity}. 

Multiwavelengh observations with VERITAS, {\it Swift}  and RXTE in 2007\cite{lsi_veritas_mw})
showed large X-ray variability with flux values typically varying between 0.5 and 
3.0$\times$10$^{-11}$ erg cm$^{-2}$ s$^{-1}$ over a single orbital cycle, but 
the TeV sampling was not dense enough to detect a correlation between the two bands.
MAGIC organized two multiwavelength campaigns. During the first campaign in 2006, 
with VLBA, MERLIN, e-EVN and Chandra\cite{lsi_magic_mw}, the gamma-ray and radio 
bands were found not to be correlated. The radio interferometers confirmed that the 
shape of the extended emission changes with the orbital period. No VHE/X-ray correlation 
could be established. In 2007, MAGIC followed the source along with the X-ray detectors 
XMM and {\it Swift} \cite{lsi_xray_correlation}
for most of a single orbit. Around the maximum of the VHE emission ($\phi$ = 0.6-0.7), 
the energy flux in X-rays was measured to be about two times larger than
the energy flux in VHE, and, as can be seen in figure \ref{fig:lsi_xray_vhe},
both bands are significantly correlated: the correlation factor is r = 0.81, 
corresponding to a random probabilty $\sim$ 5$\times$10$^{-3}$. 

The phenomenology of the source in VHE has become even more complex after
the latest observations of VERITAS in the 2008/09 and 2009/10 seasons\cite{holder,aliu}. 
These are actually
the only reported observations after {\it Fermi} started operations in 2008.
LS I +61 303 has not been detected for any of the orbital phases. In fact
making use of the full VERITAS array with nominal sensitivity has allowed to set
stringent upper limits at the level of 2\% Crab for the phase of the maximum VHE emission.
The drop in VHE emission may be correlated with the super-orbital variability of
1667 days, which, as already mentioned, is probably associated to changes in the 
circumstellar disk and may hence have an impact on the efficiency of the 
particle acceleration or $\gamma$-ray absorption processes.

\section{Uncertain VHE binaries: Cyg X-1 and HESS J0632+057}
\label{sec:2}

\subsection{Cyg X-1}

Cygnus X-1 is a High Mass X-ray Binary at 2~kpc 
distance. It represents the best established candidate for a stellar 
mass black hole, with a mass of more than 13~M$_{\odot}$.
Its optical companion is an O9.7 super-giant with a mass of 30~M$_{\odot}$ and 
a strong stellar wind\cite{cygx1_orbital}. 
The orbit is low-eccentric with radius 0.2~A.U. and 5.6 
days period.

The source belongs to the microquasar class because it displays a single-sided jet 
resolved at milli-arcsec scales with VLBA during the X-ray hard state. 
The jet's opening angle is less than 2$^{\circ}$ and the bulk velocity 
is $>$0.6c. Some authors\cite{Romero2002} have suggested that Cygnus X-1 is a
``microblazar'', where the jet axis is roughly aligned with the line of
sight. 
 
1.4 GHz radio observations show a 5 pc (8 arcmin) diameter ring structure of 
bremsstrahlung emitting ionized gas at the shock between the jet and the interstellar medium. 
The power released by the (dark) jet is of the same order 
or the bolometric X-ray luminosity and two orders of magnitude higher than that 
inferred from the radio spectrum\cite{Gallo2005}.

The results from observations in the soft $\gamma$-ray range with COMPTEL
\cite{Mcconnell2002} and INTEGRAL \cite{Cadolle2006} strongly
suggest the presence of a non-thermal component extending beyond the hard X-ray band. 
In addition, fast episodes of flux variation by a factor between 3 and 30
have been detected at different time scales, ranging from milliseconds
in the 3-30 keV band \cite{Gierlinski2003} to several hours in the
15-300 keV band \cite{Golenetskii2003}.

\begin{figure}[ht]
\begin{center}
\includegraphics[scale=.60]{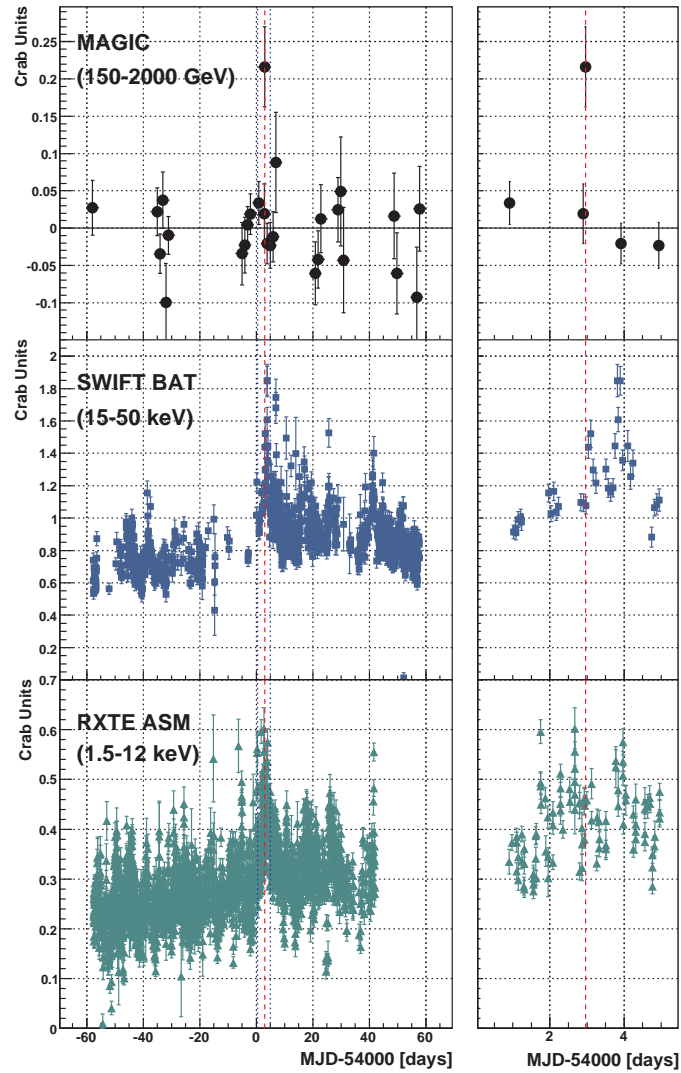}
%
%
\caption{
From top to bottom: MAGIC, {\it Swift}/BAT and
{\it RXTE}/ASM measured fluxes from Cygnus X-1 as a function of the time. 
The left panels show the whole time spanned by MAGIC observations. The
vertical, dotted blue lines delimit the range zoomed in the right
panels. The vertical red line marks the time of the MAGIC
signal. From \cite{cygx1}.}
\label{fig:cygx1}     
\end{center}
\end{figure}
MAGIC observed this microquasar for 40 h in 2006\cite{cygx1}. No VHE emission was
found either at the microquasar or at the interaction point between the jet and
the interstellar medium. Cyg X-1 however showed evidence for emission
for around 1 hour in September 24th with a pre-trial significance of 4.9$\sigma$. Given
the number of observed hours, this corresponds to a post-trial significance
of 4.1$\sigma$. The observation stopped at sunset and could only be resumed on the
following night, so the high level of VHE emission may have extended for as long as
one day.

During the 79 minutes when the VHE signal was found (MJD 54002.928 and 54002.987) 
the VHE spectrum could be
fitted to a power law with a rather soft spectral index of 3.2$\pm$0.6
and a differential flux of roughly 10\% Crab at 1 TeV.

Figure \ref{fig:cygx1} shows the VHE, soft X-ray (RXTE/ASM) and hard X-ray ({\it Swift}/BAT)
light curves of the source during the MAGIC observation campaign in 2006.
It is especially suggestive that the VHE signal was correlated with an increase
both in soft and hard X-rays, although it must be said that Cyg X-1 
was in a similarly high level of X-ray emission in the following night 
and MAGIC did not detect it in VHE. INTEGRAL also reports a historically high flux 
around the same time ($\sim$1.5 Crab between 20–40 keV and $\sim$1.8 Crab between 40–80 keV).

The long exposure also allowed to set stringent upper limits to the low-hard state of the
binary system: any steady VHE flux due to the persistent jet associated to this X-ray state
is below the present IACT’s sensitivity. 

For the time being, this remains the only evidence for VHE emission in an accreting binary system.
Detection of such an object would allow to determine the maximum particle energy which 
can be achieved in the jet of a stellar-mass black hole, a fact which may help to
understand the general mechanism of jet acceleration.

\subsection{HESS J0632+057}

HESS J0632+057 is one of very few point-like ($<$2 arcmin) unidentified sources in 
the H.E.S.S. galactic plane survey\cite{hess_gps,0632_discovery}. 
It was discovered using 13.5 hours of data collected between 
March 2004 and March 2006. The source is located in 
the region of the apparent interaction between the Monoceros Loop and the Rosette Nebula. 
Nevertheless no evidence has been found for an associated molecular cloud in  
NANTEN CO data. No object at the position of HESS J0632+057 is listed in the 
{\it Fermi}/LAT first source catalogue\cite{fermi_1fgl,he_review}.

\begin{figure}[ht]
\begin{center}
\includegraphics[scale=.50]{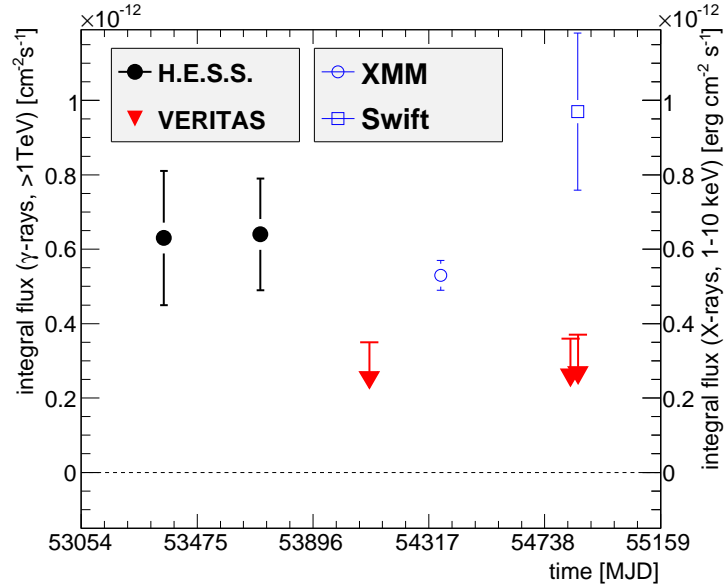}
%
%
\caption{
Light curve above 1 TeV from HESS J0632+057 assuming a spectral
shape of $dN/dE \propto E^{-\Gamma}$ with $\Gamma=2.5$. 
The downwards pointing arrows show the 99\% confidence limits derived 
from the VERITAS data. The H.E.S.S. fluxes are taken from \cite{0632_discovery}.
The X-ray fluxes measured by {\it XMM-Newton} and {\it Swift}  are indicated by open 
symbols. From \cite{0632_veritas}.
}
\label{fig:hess0632}     
\end{center}
\end{figure}
Subsequent XMM observations\cite{0632_xmm} revealed an 
X-ray source coincident with the massive B0pe star MWC 148 (HD 259440). 
However it is not known if MWC 148 is isolated or is part of a binary system.
The X-ray spectrum can be fitted to an absorbed power law with $\Gamma$=1.26 and 
displays significant variability on hour timescales. The spectral energy distribution (SED) 
of the source resembles that of LS I+61 303.

Adding to the evidence that HESS J0632+057 is associated to a VHE binary, 
VERITAS observations in 2006 and 2008/09 showed no VHE 
emission from the source\cite{0632_veritas}. The corresponding upper limits, 
shown in Fig. \ref{fig:hess0632} along with the fluxes measured by H.E.S.S., allow
to conclude that HESS J0632+057 is a variable VHE source with 99.993\% probability. 

The VERITAS campaign was simultaneous with radio and X-ray observations.
VLA and GMRT radio observations\cite{0632_radio} 
find a point-like ($<$2 arcsec) counterpart consistent with the SED and showing 
variability faster than one month. X-ray observations using {\it Swift}\cite{0632_swift} 
from MJD 54857 to 54965 also showed variability in flux (by a factor 1.8) and spectral 
shape (see Fig. \ref{fig:hess0632}). However, periodicity, 
which would strongly argue in favor of a binary system, could not be found 
on the probed timescales.

\section{Searches for other VHE binaries}
\label{sec:3}

IACTs have searched unsuccessfully for VHE emission from numerous other binary
systems over the past 20 years. We will mention here a few 
of these results. 

SS~433 is the first stellar object in which relativistic jets 
were discovered. It is an eclipsing binary system containing a 9~M$_{\odot}$ 
black hole orbiting every 13.1 days a 30~M$_{\odot}$ A3-7 supergiant star 
in a circular orbit\cite{ss433_orbital1,ss433_orbital2}. 
The system displays relativistic jets at a velocity 
of 0.26c. The jets precesses with a period of 162.4 days. SS~433 is surrounded by the radio 
shell of W50, a large 2$^{\circ}\times$1$^{\circ}$ nebula catalogued as SNR G39.7−2.0. 
It is widely accepted that its 
present morphology resulted from interactions between the jets of SS~433 and the 
surrounding nebula. Like in Cyg X-1, the production of VHE $\gamma$ rays in the SS~433/W50 
system could come both from the jet inner regions or from the interaction zones were 
the jet impacts the surrounding nebula. SS~433 has been observed by Whipple\cite{ss433_whipple}, 
HEGRA\cite{ss433_hegra}, H.E.S.S.\cite{hess_gps}, MAGIC\cite{takayuki}, VERITAS\cite{veritas_binaries}
and CANGAROO-II\cite{ss433_cangaroo}, 
but no detections have been reported either from the binary system or from the surrounding nebula.

GRS 1915+105 (or V1487 Aql) is a low mass X-ray binary consisting of a K~star with a mass 
of less than 1.3~M$_{\odot}$ on a 33~day orbit around a black hole of about 
14~M$_{\odot}$\cite{1915_orbital}.
It was established as the first galactic superluminal source after the 
discovery of two-sided radio knots moving away from the core with true velocity 
greater than 0.9c\cite{1915_mirabel}.
The source was not discovered at the HEGRA or H.E.S.S. surveys of the galactic 
plane\cite{hegra_gps,hess_gps}.
The MAGIC telescope observed GRS 1915+105 for $\sim$22 h and obtained an upper 
limit of  0.7\% of Crab Nebula flux at the 95\% confidence level\cite{takayuki}, 
above 250 GeV and assuming a power law spectrum with the photon index 
$\Gamma$=2.6. H.E.S.S. observed the object in 2004-2008 and found no evidence for a 
VHE gamma-ray signal either from the direction of the microquasar or its vicinity\cite{1915_hess}. 
An upper limit of 6.1$\times$10$^{-13}$ph cm$^{-2}$s$^{-1}$ at the 99.9\% confidence level 
was set on the photon flux above 410 GeV, equivalent to a VHE luminosity of 
$\sim$10$^{34}$erg s$^{-1}$ for a distance of 11 kpc. For this source the upper limit to
the VHE to X-ray luminosity ratio is at least four orders of magnitude lower than the ratio 
observed in the established VHE binaries. The VHE radiative efficiency of the 
compact jet is less than 0.01\% based on its estimated total power of 10$^{38}$erg s$^{-1}$.

Since VHE emission is due to particle acceleration in stellar wind/pulsar wind interaction
shock in at least one VHE binary, we may also expect to detect VHE $\gamma$-rays from
binaries where one of the components is a Wolf-Rayet (WR). MAGIC has observed 
two archetypical WR binaries\cite{wolf_rayet}: WR 147 (for 30 hours in 2007) and WR 146 
(for 45 hours between 2005 and 2007). No signal was found in any of the them. 
95\% confidence level upper limits could be set at a level of 1.5\% of the Crab flux 
for WR 147 and 5\% of the Crab flux for WR 146, both above an energy of 80 GeV.
Upper limits at higher energies up to 1 TeV are even more restrictive.

Results of VHE observations of other X-ray binaries can be found in 
\cite{veritas_binaries,takayuki,hess_binaries1,hess_binaries2}. In what remains of 
this section we will concern ourselves with a recent report of several years of 
MAGIC observations of Cyg X-3.

\subsection{Cyg X-3}

Cyg X-3 is a bright and persistent X-ray binary.  
It lies close to the Galactic plane at a distance between 
3.4 and 9.8~kpc, probably at 7~kpc\cite{Ling2009}. 
The nature and the mass of the compact object are still 
subject of debate. Published results suggest either a neutron 
star of 1.4~M$_{\odot}$~\cite{Stark2003} or a
black hole of less than 10~M$_{\odot}$~\cite{Hanson2000}. 
The identification of its donor star as a Wolf-Rayet 
star classifies it as a high-mass X-ray 
binary. Nevertheless Cyg X-3 shows a short orbital period of 4.8~hours, 
typical of low-mass binaries, which has been inferred 
from the modulation of both the X-ray and 
infrared emissions.

The source shows two main spectral X-ray states resembling the canonical states
of black hole binaries: a Low/Hard (LH) and a High/Soft (HS)
state \cite{Zdziarski2004,Hjalmarsdotter2007}. However
the LH state displays a high-energy cut-off at  $\sim$20 keV, significantly lower 
than the $\sim$100 keV value found for black hole binaries\cite{Hjalmarsdotter2004,
Zdziarski2004}. Adding to its peculiarity, Cyg X-3 is the X-ray binary displaying the 
brightest radio emission during outbursts. It frequently exhibits huge radio 
flares\cite{Braes1972} as intense as few thousand times the quiescent emission 
level of $\sim$20 mJy at 1.5 GHz. During these outbursts, which occur 
mainly when the source is in the HS state and last for several days, Cyg X-3 
reveals the presence of collimated relativistic jets\cite{Geldzahler1983,
marti2001, Miller-Jones2004}, a fact which grants it access to the microquasar class. 

\begin{figure}[h]
\sidecaption
\includegraphics[scale=.58]{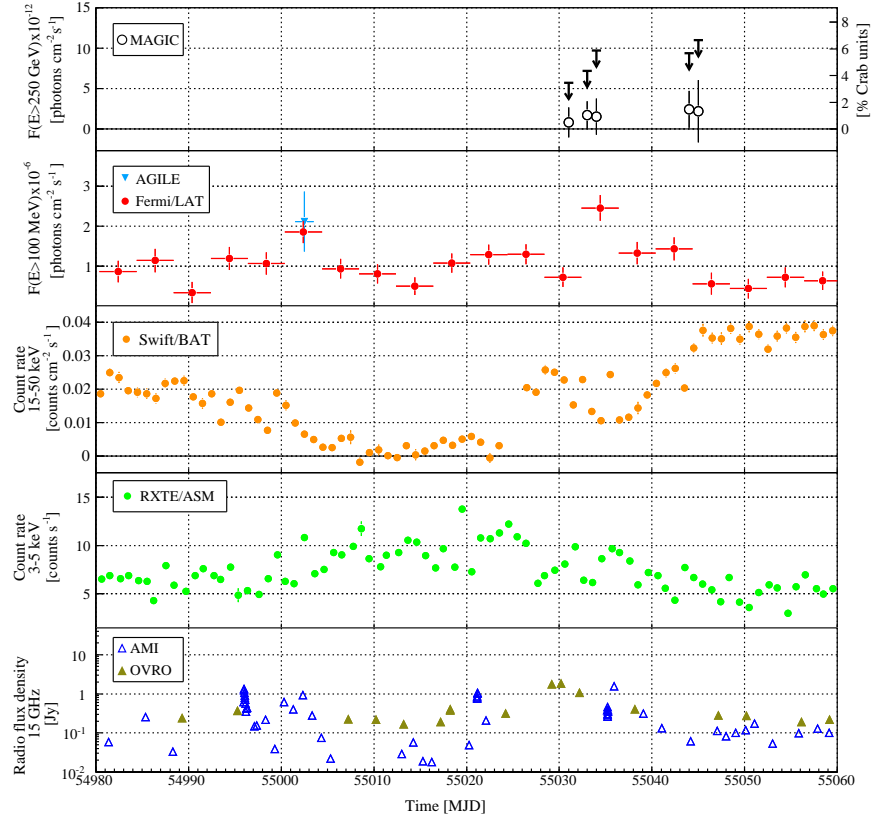}
%
%
\caption{
From top to bottom: daily light curves of Cyg X-3 measured in VHE (MAGIC, upper limits) 
at energies above 250 GeV, HE $\gamma$-rays ({\it AGILE} and {\it Fermi}/LAT), hard X-rays ({\it Swift}/BAT), 
soft X-rays (RXTE/ASM) and radio. 
From \cite{cygx3_magic}.}
\label{fig:cygx3_magic_fermi}     
\end{figure}
Cyg X-3 has also historically drawn a great deal of attention, and in fact
strongly contributed to the development of the field, due
to numerous claims of detection at TeV and PeV $\gamma$-rays. 
However, a critical analysis of these observations raised doubts on their 
validity (we refer the reader to \cite{Chardin1989} and
references within for all the reports of detection). In 
recent years, more sensitive instruments have failed to confirm those 
claims for energies above 500 GeV \cite{Shilling2001,cygx3_magic_tev2032}.
Nevertheless, the fact that this object is a microquasar with the aforementioned
strong X-ray and radio emission makes into a good candidate for VHE emission 
(see e.g. \cite{Levinson1996,Romero2003,Bosch-Ramon2006}). This radiation 
could have either an episodic nature due to the ejection of strong
radio-emitting blobs\cite{Atoyan1999} in the HS state, 
or a quasi-stationary character if it is originated in 
the persistent compact jet present during the LH state\cite{Bosch-Ramon2006}.   

The source has been very recently detected at high energy $\gamma$-rays by 
{\it AGILE}\cite{cygx3_agile} and {\it Fermi}/LAT\cite{cygx3_fermi,he_review}. {\it AGILE} found five 
$\gamma$-ray flares above 100 MeV, which were temporally 
correlated with transitional spectral states of the radio and X-ray emissions,
whereas {\it Fermi}/LAT detected an orbital modulation of the flux during periods 
of GeV high-activity which lasted for several weeks and coincided with the source
being in the HS state. {\it Fermi}/LAT has found to source to be variable with peaks
as high as $\sim$2.0$\times$10$^{-6}$ photons cm$^{-2}$ s$^{-1}$ above 100 MeV, which are 
comparable and simultaneous with the {\it AGILE} detections. The emission during the
HE active period is periodic with the orbital period of the system.

\begin{figure}[h]
\sidecaption
\includegraphics[scale=.58]{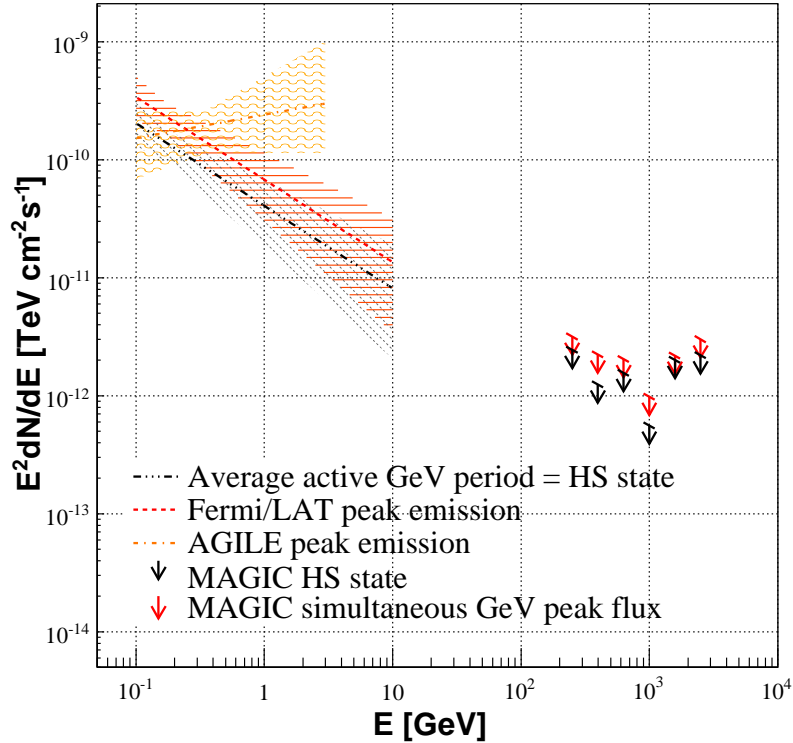}
%
%
\caption{
Cyg X-3 spectral energy distribution in the HE and VHE bands.
The lines indicate the power-law spectra derived from {\it Fermi}/LAT 
and {\it AGILE} integral fluxes and photon indices. The corresponding
errors are shown as shadowed areas.
The arrows correspond to the 95\% CL MAGIC differential flux upper limits and their slope 
indicates the assumed power-law spectrum (photon index 2.6).
The black color corresponds to the general period of {\it Fermi}/LAT 
enhanced GeV activity, which is simultaneous to the X-ray High Soft state, 
whereas, the red color corresponds to the highest HE peak (MJD 55031--55034). 
From \cite{cygx3_magic}.}
\label{fig:cygx3_SED}     
\end{figure}
MAGIC observed Cyg X-3\cite{cygx3_magic} for about 70 hours between 2006 March and 2009 August 
in different X-ray/radio spectral states and also during one of the active HE periods
reported by {\it Fermi}/LAT. No evidence was found for a VHE signal from the direction 
of the microquasar. An upper limit to the integral flux for energies higher than 
250 GeV has been set to 2.2$\times$10$^{-12}$ photons cm$^{-2}$ s$^{-1}$
(95\% confidence level). It corresponds to 1.3$\%$ of the Crab 
Nebula flux at these energies and is most stringent limit so far to the persistent 
VHE emission of this source. A search for emission was performed separately 
for each year of observations, on a daily basis and for each orbital phase. 
No significant VHE emission was found in any of the three searches (for more details
and differential flux upper limits, see \cite{cygx3_magic}). 

The VHE data 
sample was also split according to the X-ray state of the source. The upper limit to 
VHE emission during the HS state corresponds to 2.5\% of the Crab Nebula flux.
In the LH state, the VHE emission is expected to be 
produced inside the compact and persistent jets, whose total 
luminosity is estimated to be at least 10$^{37}$ erg~s$^{-1}$~\cite{Marti2005}. The MAGIC
upper limit corresponds to a VHE luminosity of 
7$\times$10$^{33}$ erg~s$^{-1}$ for a distance of 7 kpc. 
Thus, the maximum conversion efficiency of the jet power into VHE 
$\gamma$-rays is 0.07$\%$ which is similar to that of Cygnus~X-1 
for the upper limit on the VHE steady emission, but one order of 
magnitude larger than that of GRS~1915+105.

MAGIC pointed at Cyg X-3 during the second period 
of HE enhanced activity detected by {\it Fermi}/LAT in 2009, as shown 
in figure \ref{fig:cygx3_magic_fermi}.   
In particular, MAGIC carried out observations simultaneous with a 
GeV emission peak on 2009 July 21 and 22 (MJD 55033--55034), 
but did not detect VHE emission. 
The corresponding integral flux UL above 250 GeV is lower than 6\% of the Crab 
Nebula flux. 

As can be seen in figure \ref{fig:cygx3_SED},
this flux UL is roughly at the level of a power-law extrapolation 
of the HE flux measured by {\it Fermi}/LAT, assuming the photon index $\Gamma$=2.6 
which {\it Fermi}/LAT has measured for both the low and high state of the source, 
but it is significantly below the extrapolation of the {\it Fermi}/LAT flux 
but assuming the photon index measured by {\it AGILE} only during the high state 
($\Gamma$=1.8). Both results point to a cutoff in the energy spectrum
of the source at energies in the range between a few GeV and 250 GeV.

\section{Conclusions}

VHE binaries may be powered by the interaction between a pulsar wind
and the wind of its companion star, or by an accretion-driven jet. 

Three binary systems have been established at the VHE band:
PSR~B1259-63, LS 5039 and LS I +61 303. Emission in the first binary
stems from pulsar wind/stellar wind interaction, and the other two 
objects are consistent with the same scenario. All three sources 
display a complex phenomenology. Even if there is a significant 
component of the VHE emission which is associated to the orbital period,
variations from orbit to orbit are observed. LS I +61 303
significantly dropped in brightness over the past two years and recent
observations have in fact failed to detect it for any orbital phase,
a fact which may be linked to superorbital variability observed at
other wavelengths.

Its variability at VHE and lower energies, its point-like character and
its SED make HESS J0632+057 into a plausible candidate to become the
fourth VHE binary. On the other hand, a black hole X-ray binary, Cyg X-1,
has shown a marginal episodic signal at VHE, which may eventually make it
into the first VHE microquasar.

Searches for other well-known X-ray binaries have proved unfruitful.
Recent observations of Cyg X-3 in a wide variety of X-ray and radio
states and during a high energy $\gamma$-ray flare detected by {\it Fermi}/LAT
have not shown evidence for simultaneous VHE emission.

\end{document}